\documentclass[12]{iopart}

\usepackage{cite}
\usepackage{graphicx}

\begin{document}

\title[One-to-one direct modeling of experiments and astrophysical scenarios]{One-to-one direct modeling of experiments and astrophysical scenarios: pushing the envelope on kinetic plasma simulations}

\author{R A Fonseca$^{2,1}$, S F Martins$^1$, L O Silva$^1$,  J W Tonge$^3$, F S Tsung$^3$, and 
W B Mori$^3$}

\address{$^1$GoLP/Instituto de Plasmas e Fus\~ao Nuclear, Instituto Superior T\'ecnico, 1049-001 Lisboa, Portugal}
\address{$^2$Departamento Ci\^encias e Tecnologias da Informa\c{c}\~ao, Instituto Superior de Ci\^encias do Trabalho e da Empresa, 1649-026 Lisboa, Portugal}
\address{$^3$UCLA Plasma Simulation Group, University of California Los Angeles, Los Angeles CA, U.S.A.}
\ead{ricardo.fonseca@ist.utl.pt}

\begin{abstract}
There are many astrophysical and laboratory scenarios where kinetic effects play an important role. These range from astrophysical shocks and plasma shell collisions, to high intensity laser-plasma interactions, with applications to fast ignition and particle acceleration. Further understanding of these scenarios requires detailed numerical modelling, but fully relativistic kinetic codes are computationally intensive, and the goal of one-to-one direct modelling of such scenarios and direct comparison with experimental results is still difficult to achieve. In this paper we discuss the issues involved in performing kinetic plasma simulations of experiments and astrophysical scenarios, focusing on what needs to be achieved for one-to-one direct modeling, and the computational requirements involved. We focus on code efficiency and new algorithms, specifically on parallel scalability issues, namely on dynamic load balancing, and on high-order interpolation and boosted frame simulations to optimize simulation performance. We also discuss the new visualization and data mining tools required for these numerical experiments and recent simulation work illustrating these techniques is also presented.

\end{abstract}

\pacs{5265, 5238, 5240}
\submitto{\PPCF}
\maketitle

\section{Introduction}

Due to the highly nonlinear and kinetic processes that occur in many plasma physics scenarios, ranging from astrophysical shocks and plasma shell collisions \cite{Fonseca:2003p174,Silva:2003p160}, to high intensity laser-plasma interactions, with applications to fast ignition \cite{Ren:2004p703} and particle acceleration \cite{Silva:2004p709,Tsung:2006p748, Lu:2007p690}, the key tools for modeling these problems are fully relativistic three-dimensional kinetic codes \cite{Fonseca:2002p704, Pukhov:1999p1052, Sentoku:2002p1149, Nieter:2004p968, Ruhl:2002p1150, Geissler:2006p1314, Bowers:2008p904, Spitkovsky:2008p1234,Benedetti:2008p1322}. However, the goal of one-to-one direct modeling of such scenarios and direct comparison with experimental results is still difficult to achieve. These codes are computationally heavy, and until recently have only been able to model limited dimensions and short time/length scales. 

The computational requirements involved in performing these simulations are pushing the limits of particle in cell (PIC) simulations. The advent of modern high performance computing (HPC) systems, with massively parallel machines going up to $\sim 1$ PFlop/s \cite{top500}, has allowed these codes to be applied to much larger problems, but full 3D simulations of large space/time scales are still extremely difficult to achieve. If we look at the laser wakefield accelerator \cite{Tajima:1979p908}, recent scalings \cite{Lu:2007p690} indicate that to reach an energy of $10 \,\rm{GeV}$, the accelerating length must be on the order of $\sim 0.5 \,\rm{m}$, with a plasma density of $\sim 10^{17} \,\rm{cm}^{-3}$. Since the laser wavelength, $\lambda_0 \sim 1 \,\rm{\mu m}$, needs to be resolved, the simulation grid will be very fine, and the total number of iterations required will be on the order of $\sim 10^7$. Another relevant scenario is that of relativistic shocks, that are relevant in many astrophysical \cite{Silva:2003p160} and laboratory scenarios \cite{Silva:2004p709}. These scenarios also present an extremely challenging task, since they require simultaneously modeling very different time and length scales. The simulation must be run for much longer than the time it takes the shock to form ($\sim 100 \,\omega_p^{-1}$ for the heavier species) but $\omega_p$ for the electrons must also be resolved. The spatial extent of the simulation must model both shock formation and propagation at relativistic velocities, while simultaneously resolving the electron plasma skin depth. These numerical experiments require total number of particles of the order $\sim 10^{10}$ to be followed for $\sim 10^6 - 10^7$ timesteps, with total memory requirements going up to $\sim 1 \,\rm{TB}$. 

To achieve the goal of one-to-one modeling of these scenarios existing numerical frameworks must therefore be extended. For realistic simulation parameters the total number of CPU hours involved can be up to $\sim 10^6$ hours, so sustained efficient use of large parallel HPC systems is required. This is a very challenging task, as it requires parallel scalability for thousands CPUs, specifically ensuring that the computational load is evenly distributed among nodes \cite{FERRARO:1993p829} over the total simulation duration. Concurrently, it is also necessary to further optimize simulation models, taking full advantage of used CPU cycles, and insuring simulation stability for such large iteration numbers. Finally, we should also focus on the type of diagnostics that these simulations require. The amount of data produced increases significantly and sophisticated more intelligent diagnostics are required to extract relevant information. Besides standard diagnostics, detailed evolution of particles in phase-space is an essential tool, that would greatly improve our understanding of the underlying physics, since some of the most relevant phenomena (e.g. particle acceleration) are associated with the dynamics of a small subset of the total number of particles in the simulation.

In this paper we discuss the issues involved in performing full scale numerical experiments of astrophysical and laboratory scenarios. The article is organized as follows. First we present the dynamic load balancing algorithm to improve scaling on high-end parallel machines. Next we describe the required improvements to the simulation algorithm, in terms of reference frame choice and high order particle shapes, extending the temporal and spatial scales that can modeled with these codes. Section 4 discusses advanced visualization and diagnostics, and presents recent simulation work illustrating the techniques discussed in previous sections. Concluding remarks and a summary are offered in Section 5.

\section{Scaling for high-end HPC systems}

The electromagnetic PIC algorithm \cite{Dawson:1983p1151} uses a particle-mesh technique to model a sample of the full 7D phase-space evolution of a kinetic plasma. In this algorithm the full set of Maxwell's equations are solved on a grid using the electrical current and charge density calculated by weighting discrete particles onto the grid. Each particle is then pushed to a new position and momentum via the self-consistently calculated fields. This algorithm is well suited for parallelization, since all operations can be made locally, i.e., not requiring information from the whole simulation box. This allows for parallelization based on spatial partitions with small communication only with neighboring nodes, with great parallel efficiency. However, as the number of parallel nodes used increases, other issues need to be considered. One critical issue when scaling for massively parallel HPC systems is that of load balancing; for the parallel efficiency to be high it is essential to divide the workload evenly across the thousands of CPUs in use, as simulation performance is generally dictated by the CPU with the most load. In PIC codes the parallel spatial partitions can be chosen so that initially the computational load is evenly distributed. However, as the simulation progresses, particles are free to move between parallel partitions, which may result in a large load unbalance, and a catastrophic performance hit. 

There are many scenarios, such as cluster explosions \cite{Peano:2005p175}, ion acceleration \cite{Silva:2004p709} or full scale fast ignition simulations \cite{Ren:2004p703}, where we must also model a large vacuum region surrounding the plasma. For example, in the case of ion acceleration \cite{Silva:2004p709}, the plasma uses less than 20\% of the simulation box, which would mean that for a regular one dimensional parallel partition over 80\% of CPUs would have no particles to push. A choice of an irregular partition could solve the problem initially but it would not be enough; for example, in simulations of cluster explosions where a small plasma sphere explodes as a result of laser interaction, most simulation particles will initially be very close to the center of the simulation box, and then expand to fill the whole simulation space. If parallel boundaries are maintained at fixed positions, the computational load will be concentrated on a few parallel nodes, and overall efficiency will be very low. 

The solution to this problem is to have the code dynamically readjust the parallel partitions in response to the changes in particle positions \cite{FERRARO:1993p829}. To achieve this the code must build a computational load estimate from the current simulation state, taking into account the computational weight for pushing particles and solving field equations on the grid. Using this information the code can now decide what is the optimal choice of parallel partition for the current simulation state, and reshape the simulation accordingly. This is done along a single chosen direction; if any parallel boundaries along other directions exist they are kept fixed. The smallest domain that can be chosen is limited by the particle interpolation scheme used and is on the order of a few cells. In this limit the ratio between computation (number of particles/cells handled by the node) and communication (number of particles crossing node boundaries and number of cells exchanged with neighbor nodes) will become lower. However, with present communication networks, this poses a negligible overhead, and the code can safely use the smallest possible domains. In some critically unbalanced scenarios this may lead to volume ratios between largest and smallest domains as high as two orders of magnitude, but for typical simulations this value is on the order of $\sim 5$.

\begin{figure}
\begin{center}
\includegraphics{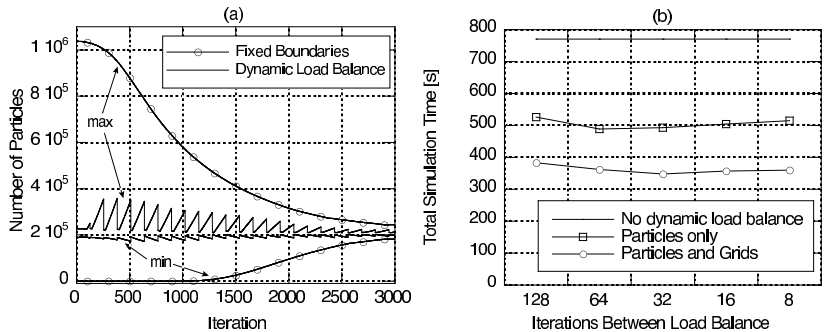}
\caption{(a) Shows the maximum (max) and minimum (min) number of particles in each parallel node for fixed boundary and dynamic load balancing runs. (b) Shows the simulation time versus dynamic load balance frequency. }
\label{fig_dynlb}
\end{center}
\end{figure}

Finally, the code should repeat this procedure at some interval to maintain a good average load balance between all simulation nodes. This can be a lengthy procedure, so for the benefits to overcome the costs it should not be triggered too often. Figure \ref{fig_dynlb} (a) shows a dynamic load balancing algorithm in progress for a test simulation. At every 128 timesteps the algorithm rearranges the simulation boundaries in order to maintain an even number of particles on every node. Figure \ref{fig_dynlb} (b) shows the impact on performance of this technique as a function of the frequency at which dynamic load is triggered. As we can see, there is a two fold performance improvement. For optimal results, it is crucial that the grid calculations are included when estimating the computational load, and not just the total number of particles. We also see that load balancing should not be done too often as it will slightly penalize overall performance. As a rule of thumb, choosing a dynamic load balance frequency of $\sim 100$ iterations generally yields good results. Using this technique we were able to obtain a speedup of a factor of $\sim 4$ for the above mentioned cluster explosion simulations \cite{Peano:2005p175}.

\section{Improving the Simulation Algorithm}

In many relevant situations the simulation algorithm needs to model objects crossing at relativistic velocities. This is the case of high intensity laser/particle beam plasma interaction, where a generally short beam is propagated over a long plasma length. In PIC codes these simulations are generally done with a so called ``moving window'' algorithm \cite{Tzeng:1996p898}. This means that  the simulation is run in the laboratory reference frame, but the simulation window is advanced at the speed of light, with all simulation data being shifted accordingly at every time step. For small laser propagation distances, on the order of a few Rayleigh lengths this has proven to be a very useful technique. However, if we were to use this technique to a $10 \,\rm{GeV}$ laser wakefield accelerator with the parameters described in the introduction, we would still require on the order of  $\sim 10^7$ iterations and $\sim 10^7$ cpu hours, that would take $\sim 8$ months to run on 1024 CPUs. Another example is that of astrophysical shocks. In these scenarios the simulation must follow the shock formation and propagation at relativistic velocities over long distances. For a full scale 3D kinetic simulation with realistic mass ratios we would require $\sim 10^7$ timesteps and a similar simulation time to the previous example.

Recent work with charged particle beams has shown \cite{Vay:2007p134} that performing these simulations in a boosted frame, where the laser/particle beam/shock front is close to rest, may be advantageous and significant performance gains (of up to 2 orders of magnitude) can be achieved. This reference frame optimizes the simulation geometry/duration, given that the system is more compact spatially, the interaction time is shorter, and the ratio between largest to smallest time scale is minimized. The initialization of a simulation in a boosted reference frame for the initial time step is straightforward: it is only necessary to convert the plasma density and velocity distribution, and initial electromagnetic fields (e.g. laser pulse) using a standard Lorentz transformation. However, as the simulation progresses, fresh plasma must be injected from the front wall of the simulation, moving backwards with the same speed as the reference frame. It will also be necessary to include ions in the simulation (or their equivalent neutralizing current) since the background electrons now inject an electrical current of $J_x' = - c \rho \sqrt{\gamma^2 - 1}$ into the simulation box, where $\rho$ is the electron charge density in the rest frame. The overhead of modeling background ions (which is generally unnecessary in laboratory frames) can be minimized if these are considered to be free streaming, which effectively corresponds to having a fixed ion background in the rest frame. 

The PIC algorithm needs no change since the model is invariant under a Lorentz transformation. However, running the simulation in this frame poses special difficulties, and special caution must be taken in the choice of boosted frame. Since the background plasma is moving at relativistic velocities, any roundoff error when adding the current of background electrons and ions will generally lead to a numerical electromagnetic instability, seeded from the numerical noise. There are also factors that limit the Lorentz factor of the boosted frame (and hence the performance gain). The spatial compression along the transformation direction limits the spatial resolution, especially for backward moving radiation, and a balance must found to retain all the relevant physical scales as well as to prevent other electromagnetic instabilities (e.g. \cite{Dieckmann:2006p1366}) from growing. The FTDT field solver used in this algorithm will generally underestimate the phase-velocity of high frequency electromagnetic modes, making it less than the speed of light. This leads to numerical \v{C}erenkov radiation, which is especially critical since all plasma particles are moving relativistically in the boosted frame. This can be compensated through  finer grid/time step, which would remove the performance benefits. Alternatively special field solvers \cite{Greenwood:2004p136} that compensate the lower phase-space velocities at high frequencies, or low pass-filtering in the spatial (k)-wave spectrum could be used. We tested both solutions and have found the latter to be the most efficient if filtering is kept to a minimum as it may also eliminate small scale structures in the simulation. 

\begin{figure}
\begin{center}
\includegraphics{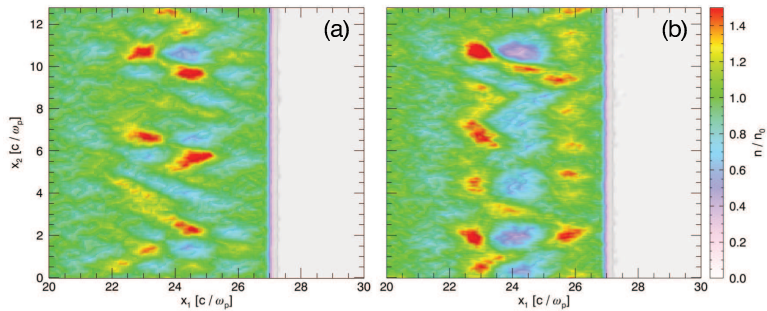}
\caption{Plasma shell collisions simulated in a rest frame (a), and results from a boosted frame reconstructed on the rest frame (b).}
\label{fig_boost}
\end{center}
\end{figure}

In this reference frame relativity poses one final problem: for the simulation results to be easily analyzed and compared with standard rest frame simulations they need to be converted back to this frame. Besides performing the standard Lorentz transforms between Lorentz reference frames, one needs also to take into account that events that are simultaneous in one of the frames will happen at different times in the other. For example, if the simulation is performed in a frame moving with velocity $v$ and corresponding Lorentz factor $\gamma$ building a diagnostic of charge density $\rho$ in the rest frame for a given time and spatial region, $ \rho = \gamma( \rho' + v J_x' / c^2)$, requires both the charge density $\rho'$ and electrical current $J_x'$ in the boosted frame for the time $t'$ and position $x'$ corresponding to the same time $t$ and position $x$ in the rest frame, $\{x' = \gamma(x-vt), t' = \gamma(t - v x / c^2)  \}$. This means that for a fixed time in the rest frame each point $x$ will require information from different boosted times $t'$. 

Figure \ref{fig_boost} compares the results from 2D simulation of the collision of two plasma shells performed in different reference frames, showing filamentation due to the Weibel instability. The boosted simulation results shown (\ref{fig_boost}b) were converted into the laboratory reference frame for comparison with simulation (\ref{fig_boost}a) results in this frame. The two simulations show qualitatively the same results, with same filament size and growth rate. The differences are the result from the initial microstate of both simulations, since this instability grows from noise. Regarding the performance boost, we have recently concluded full scale simulations for the previously mentioned 10 GeV accelerator scenario  (to be published). We were able to obtain a performance boost of $\sim 100$ using this technique, completing the simulation of the laser propagation over $\sim 0.5 \,\rm{m}$ in less than a week, with excellent agreement with the theoretical models \cite{Lu:2007p690}.

There are however other scenarios where this technique cannot be employed. For example, in fast ignition scenarios a high intensity $\lambda \sim 1 \,\rm{\mu m}$ laser pulse will have to be propagated over $\sim 100 \,\rm{\mu m}$ of plasma with densities ramping up from subcritical to as much as $\sim 10^5$ times critical. Most of the fast ignition target needs to be modelled and $\sim 10^6 - 10^7$ timesteps will be required. To model these very long problems, it is also critical to insure that inherent PIC algorithm issues do not affect the overall results,  such as energy conservation and numerical self-heating, a well known phenomenon related to the numerical interaction between the simulation particles and grid \cite{Hockney:1971p912}. In the PIC algorithm this interaction is done through particle weighting. In this process, the particles are assumed to represent a charged cloud with a given spatial charge distribution, generally referred to as particle shape. For each particle, the charge and electrical current on the grid are calculated as the deposition of this charge distribution, and particle forces are calculated as the integral of the forces over these clouds. For simulations with up to $\sim 10^4$ timesteps using linear particle shapes \cite{Villasenor:1992p132} yields good results. However, these shapes have a discontinuous derivative, and a large number of particles per cell must be used to reduce numerical noise for longer runs. Although computationally heavy, this has the advantage of increasing the resolution of the phase-space of the simulations, and enabling us to probe very fine details of the physical system being modeled. Alternatively, if the necessary resolution is already achieved with less particles per cell, we can use a higher order interpolation scheme  \cite{Esirkepov:2001p133}. In this case the implemented particle shape will be ``smooth'', i.e., with no discontinuities in its derivative, and the noise and energy conservation properties of our algorithm can be further improved. 

\begin{figure}
\begin{center}
\includegraphics{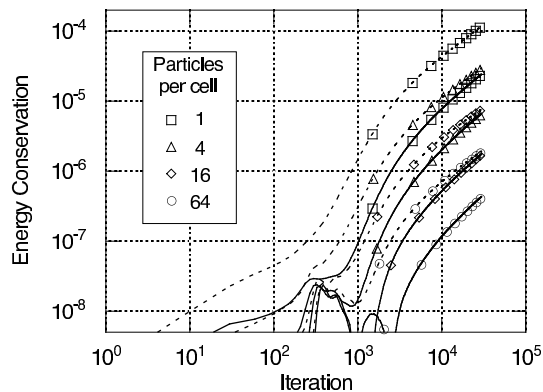}
\caption{Comparison of energy conservation for linear (dashed line) and quadratic (solid line) particle interpolation for several choices of number of particles per cell.}
\label{fig_spline}
\end{center}
\end{figure}

This will require more operations to be performed for each particle (in 3D quadratic interpolation requires $\sim 4$ times more operations than linear interpolation) but the end result can be a faster overall simulation, since the same noise level and energy conservation can be achieved with fewer particles. Figure \ref{fig_spline} shows energy conservation on 2D simulation of colliding electron-positron shells in astrophysical scenarios \cite{Fonseca:2003p174}. For the same number of particles per cell, the energy conservation using quadratic interpolation is generally superior by about 1 order of magnitude. It is also clear that the same level of energy conservation can be achieved with fewer particles in the quadratic interpolation case. The simulation with only 1 particle per cell with quadratic interpolation has better results than the simulation with 4 particles per cell with linear interpolation. In this case, given the lower number of total simulation particles, the quadratic interpolation run is almost twice as fast as the linear interpolation run. This is also true in three dimensions, where an increase in the number of particles per cell has an even bigger impact on the total simulation size.

\section{Advanced Visualization and Diagnostics}

As simulations approach one-to-one direct modeling of experiments and astrophysical scenarios, the data being produced is becoming both very large and extremely complex. Sophisticated diagnostics are therefore required to extract and present the simulation results in a manner that is easily accessible and understandable, and enabling scientists to gain physical insight into the relevant processes. The visualization routines must not only be able to display grid and particle data, but must also allow for the combination of multiple datasets, datamining and analysis, and to observe in detail the time evolution of given simulation quantities, and must do so in a manner that is easily accessible and understandable. The visualization process is generally overlooked when discussing high end numerical experiments. However, it is critical that it does not become the bottleneck for simulation output. The visualization process should be as simple as possible, requiring a minimum of user intervention. This means that a tight integration between simulation and visualization is essential, including all relevant metadata into diagnostic files, and allowing for the straightforward creation of presentation quality plots. 

\begin{figure}
\begin{center}
\includegraphics{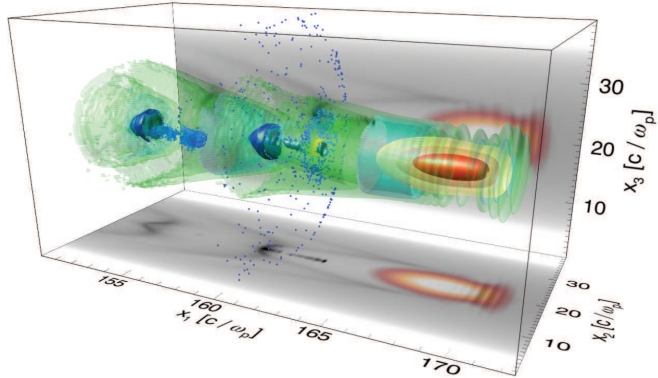}
\caption{Laser Wakefield accelerator simulation showing laser field, plasma wake, and accelerated particles.}
\label{fig_vis}
\end{center}
\end{figure}

\begin{figure}
\begin{center}
\includegraphics{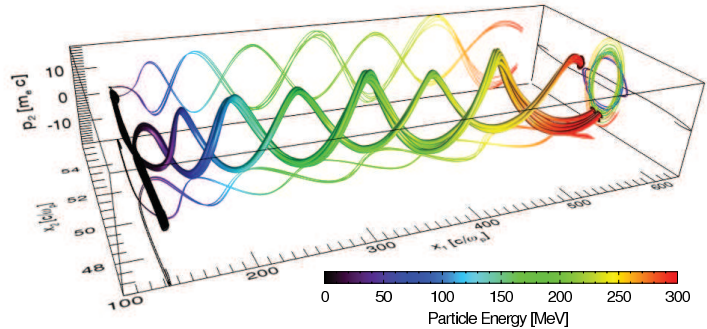}
\caption{Phase-space evolution of accelerated particles in a Laser-Plasma wakefield accelerator simulation showing betatron oscillation orbits.}
\label{fig_tracks}
\end{center}
\end{figure}

One situation where visualization plays a critical role is the simulation of laser plasma accelerator experiments \cite{Leemans:2006p1368}, specifically where electron self-injection plays an important role. A complete and detailed understanding of the self-injection mechanism is essential for controlling these accelerators. However, extracting the necessary physical insight from these simulations is not straightforward, as the number of simulation particles that is self injected for acceleration is only $\sim 10^4$ out of the total $\sim 10^9$ particles in the simulation. Furthermore, the exact details of the injection must be known, requiring detailed time history of these particles and the fields that they interact with. Figure \ref{fig_vis} shows an example of the typically required visualizations in these scenarios. This specific example combines two grid datasets and select particle data. Only a subset of the available particles is displayed (specifically particles with energies above 20 MeV). Another relevant diagnostic is that of particle tracking, that allows the observation of the detailed evolution in a 7D (space, momenta and time) phase-space of a selected group of simulation particles. If a high time resolution is required, storing diagnostic information for the complete set of particles in the simulation quickly becomes impossible, given the large storage requirement (a single dump of $10^9$ particles requires $\sim 50 \,\rm{Gb}$). The solution involves running the simulation twice: once to perform data mining and to find out what the interesting particles are and a second time to write tracking information about these particles. Figure \ref{fig_tracks} shows an example of this diagnostic applied to a Laser-Plasma wakefield accelerator simulation. The 10 particles with the highest energy at the end of the simulation were chosen, and the simulation was run again to store tracking information for every time step.

\section{Conclusions}

With the HPC systems that are now coming online, the unprecedented computational power available opens the door to very large scale numerical simulations. Together with the changes described in this paper, these put the goal of one-to-one kinetic plasma simulations within reach. The techniques presented here make better use of large scale parallel HPC systems, optimize the simulation algorithm and geometry, and give us the necessary tools to extract the necessary physical insight from the data being produced. With these tools large time and spatial scales can finally be successfully modeled using fully relativistic kinetic codes.

The developments presented this paper were done in the OSIRIS \cite{Fonseca:2002p704} framework, which is, to the best of our knowledge, the only code currently supporting all of these features. Other codes also implement many of these features; for example \cite{Pukhov:1999p1052}Ê supports dynamic load balancing, \cite{Nieter:2004p968,Benedetti:2008p1322} support boosted frames, \cite{Nieter:2004p968,Benedetti:2008p1322,Spitkovsky:2008p1234} support higher order interpolation, and \cite{Geissler:2006p1314} supports particle tracking. Other possibilities on improving the PIC algorithm and expanding its validity are also being pursued, like automatic mesh refinement \cite{Vay:2004p1369} and additional physics \cite{Sentoku:2008p1328,Ruhl:2002p1150}. Finally, a recent overview and benchmark of simulation codes used in the modeling of plasma accelerator experiments can be found in \cite{CGRGeddes:2008p1370}.

\ack

This work partially has been partially supported by FCT (Portugal) under grants PDCT/FP/63928/2005, PDCT/FIS/63915/2005, and POCI/66823/2006 and by the European Community - New and Emerging Science and Technology Activity under the FP6 ``Structuring the European Research Area'' program (project EuroLEAP, contract number 028514). Some of the simulation work presented here was produced using the IST Cluster (IST/Portugal). The authors would also like to thank Dr. Viktor Decyk for helpful discussions.

\section*{References}

%\bibliographystyle{iopart-num}
%\bibliography{fonseca}

\providecommand{\newblock}{}

\end{document}